\documentclass[prd,preprint,nofootinbib]{revtex4}
\usepackage{amssymb}
\usepackage{epsfig}

\begin{document}
\title{ 
Reheating and gravitino production in braneworld inflation
}

\author{Edmund J. Copeland}
\email{Ed.Copeland@nottingham.ac.uk}
 \affiliation{
 School of Physics and Astronomy, University of Nottingham, 
 University Park, Nottingham NG7 2RD, United Kingdom
 }

\author{Osamu Seto}
\email{O.Seto@sussex.ac.uk}
 \affiliation{
 Department of Physics and Astronomy, University of Sussex, 
 Brighton BN1 9QJ, United Kingdom
}


\begin{abstract}
We consider the constraints that can be imposed on a wide class of Inflation
models in modified gravity scenarios in which the Friedmann equation is 
modified by the inclusion of $\rho^2$ terms,
where $\rho$ is the total energy density.  
In particular we obtain the reheating temperature and gravitino abundance
associated with the end of inflation. Whereas models of chaotic inflation and 
natural inflation can easily avoid the conventional gravitino 
overproduction problem,
we show that supersymmetric hybrid inflation models (driven by both F and
D-terms) do not work in the $\rho^2$ dominated era. 
We also study
inflation driven by exponetial potentials in this modified background, and
show that the gravitino production is suppressed enough to avoid there being a
problem, although other conditions severely constrain these models. 
\end{abstract}

\pacs{}


\vspace*{3cm}
\maketitle


\section{Introduction}

Inflation has proved to be a successful paradigm within which to explain a
number of observational features of our Universe. However, any successful model
must satisfy a number of physically motivated constraints. One of these is
known as the ``gravitino problem" \cite{KhlopovLinde, KKM}, 
in which the gravitinos produced at the end of a period of inflation decay 
after Big Bang Nucleosynthesis (BBN). 
Unless the gravitino is the lightest supersymmetric particle (LSP) or very heavy,
then its energetic daughters would cause the destruction of the light nuclei, 
thereby destroying the successful prediction of BBN. 
In order to avoid this problem, an upper bound must be imposed on the
reheating temperature after inflation, a bound which depends on the mass of the
gravitino and the efficiency of the hadronic processes  \cite{KKM}.

Recently, a novel solution to the gravitino problem was proposed in the context
of modified gravity models in Ref.~\cite{OkadaSeto}. It takes advantage of some
important cosmological features arising in a class of brane world models,
where our four dimensional universe is realized on 
 the ``3-brane'' located at the boundary of the bulk spacetime (for a review
see \cite{LangloisReview}). 
One particularly simple model was proposed by Randall and Sundrum \cite{RS}.
The action of the model is given as
\begin{equation}
S = \frac{M_5^3}{2}\int_{\mathcal{M}_5} d^5x \sqrt{-g}(R-2\Lambda_{\rm bulk})
 +\int_{\mathcal{M}_4}d^4x \sqrt{-g}(-\lambda+\mathcal{L}_{matter}),
 \label{Action}
\end{equation}
 where $M_5$ is the five dimensional Planck mass, 
 $\Lambda_{\rm bulk}$ is the bulk negative cosmological constant, 
 $\lambda$ is the tension of ``3-brane" and $\mathcal{L}_{matter}$ denotes 
 the lagrangian for the matter on the brane. In this paper, we will be
 considering bounds on the gravitino, hence ideally we would want to consider
 the supersymmetric generalization of the Randall Sundrum (RS) model
 \cite{SusyRS}. However, we do not have to go that far, because the modified
 Friedmann equation arising in these models is the same for both cases, as it
 arises from the bosonic part of the action. 
The Friedmann equation for a spatially flat four dimensional
 spacetime is found to be \cite{Shiromizu:1999wj,Binetruy:1999ut}
\begin{equation}
H^2 = \frac{1}{3M_P^2}\rho\left(1+\frac{\rho}{2\lambda}\right)
 \label{BranefulFreidmann}
\end{equation}
 where $H$ is the Hubble parameter, 
 $M_P \simeq 2.4\times 10^{18}$ GeV is the reduced Plank mass, 
 which is related to $M_5$ and $\lambda$ through
\begin{equation}
M_P^2 = \frac{6M_5^6}{\lambda} ,
\label{M_5-M_P}
\end{equation}
 and $\rho$ is the total energy density of matter on the brane.

Turning our attention to gravitino production at the end of a period of
inflation, we can write down the Boltzmann equation for this process as
\begin{equation}
\frac{d n_{3/2}}{d t}+3H n_{3/2} = 
 \langle\sigma_{tot} v\rangle n_{rad}^2,
\label{Boltz:n}
\end{equation}
where $n_{3/2}$ and $n_{rad}$ are the number density of gravitino and
relativistic particles respectively, and $\langle\sigma_{tot} v\rangle$ is the
thermal averaged product of the interaction cross section and relative velocity
of the interacting particles.
Introducing $Y_{3/2} \equiv n_{3/2}/s$ where $s$ is the entropy density, the
Boltzmann equation can be rewritten in terms of the temperature $T$ as
\begin{equation}
\frac{d Y_{3/2}}{d T} = 
 -\frac{s\langle\sigma_{tot} v\rangle}{H T} Y_{rad}^2 .
 \label{GravitinoBoltamann}
\end{equation}
The final abundance of the gravitinos can be estimated by integrating 
 Eq.~(\ref{GravitinoBoltamann}) from the reheating temperature $T_R$ 
 to a lower final temperature $T_{\rm low}$.
In particular, since $s \sim T^3$ and $H \propto T^2$ in standard cosmology, it
follows that $Y_{3/2} \propto T_R$. Hence we obtain an upper bound on the
reheating temperature, by constraining the abundance of gravitinos.
To put it concretely, the reheating temperature after inflation is constrained
to be
\begin{equation}
\label{TR-constraint}
T_R \lesssim 10^6 - 10^8 \textrm{GeV},
\end{equation}
depending on the gravitino mass and the efficiency of the relevant hadronic
effect \cite{KKM}.
Things change though when we consider the modified equation
(\ref{BranefulFreidmann}) arising in  brane world scenarios. In that case a
second important temperature, $T_t$ can be defined which allows us to separate
the issue of the gravitino abundance from the reheating temperature at the end
of inflation in the $\rho^2$ regime. $T_t$ is the transition temperature which
marks the epoch when the $\rho^2$ term becomes sub-dominant in the modified
Friedmann equation (\ref{BranefulFreidmann}) and is therefore given by 
\begin{eqnarray}
\frac{\pi^2}{30}g_* T_t^4 = 2\lambda = \frac{12M_5^6}{M_P^2}
\label{define-Tt}
\end{eqnarray}
where $g_*$ is the total number of degrees of freedom of the relativistic
particles. 
In Ref.~\cite{OkadaSeto} it was shown that by replacing $T_R$ in the standard
cosmology with $2 T_t$, then if $T_t$ is low enough, for example 
\begin{equation}
 T_t \lesssim 10^6 \textrm{GeV} 
 \label{Tt}
\end{equation}
 or equivalently
\begin{equation}
 M_5 \lesssim 10^{10} \textrm{GeV} , \label{M5bound}
\end{equation}
 it is then possible to separate the gravitino abundance from 
 the reheating temperature after inflation. In particular if the constraint
Eq.~(\ref{Tt}) is satisfied then even with high reheating temperatures, there is
no gravitino problem.  

In this paper, we investigate whether inflation models can benefit
through the brane world type cosmological solution to the gravitino problem.
Although there has been considerable work investigating Inflation models where 
the inflaton is confined to the brane and inflation is driven during the
$\rho^2$ dominating stage 
 \cite{MaartensWandsBassettHeard,TsujikawaLiddle,Mazumdar,Felipe}, here 
 we re-examine these inflationary braneworld models 
 in light of Ref.~\cite{OkadaSeto}.
As particular examples, we study models with quadratic, exponential and 
 cosine potentials as well as more general hybrid potentials.
Although, as is well known,
 it is a non-trivial task to construct viable
 quadratic potential models in the framework of supergravity \cite{MSYY},
 we study that example here as representing a typical example of polynomial
 inflaton potential \cite{Linde}.
We will have to investigate the exponential potential separately because of the
unique way of reheating associated with these models. 

Inflation in the context of supergravity and in a
class of M-theory inspired models is well known to be problematic, both in the
context of D and F term potential driven inflation \cite{lyth-non-renorm}.
Basically, the presence of non-renormalizable terms in the potential lead to
the famous $\eta$-problem, meaning that insufficient e-foldings of inflation
occur. In this paper we will once again have to confront this problem in the
context of inflation occurring in the $\rho^2$ dominated regime.  The paper is
organized as follows. After summarizing the basic equations in the next
section, we study a number of
inflation models except the exponential potential in Sec. III. We leave the
exponential case to Sec. IV, and conclude in Section V.

\section{Density perturbations in brane world inflation}

In a brane world where inflation is driven by the potential of the inflaton on
the brane, the Lagrangian for the canonically normalized inflaton field, $\phi$
 is given by 
\begin{equation}
 \mathcal{L}_{inflaton}(\phi) = -\frac{1}{2}(\partial\phi)^2 -V(\phi),
\end{equation}
 where, for simplicity we have omitted possible interaction terms 
 between the inflaton and other matter fields.
The inflaton equation of motion is 
\begin{equation}
 \ddot{\phi}+3H\dot{\phi}+V'(\phi)=0,
\end{equation}
where a prime denotes a derivative with respect to the inflaton field.
Now during inflation, $\phi$ obeys the slow roll equation
$3H\dot{\phi}+V'(\phi)
=0$.
We can define the slow roll parameters $\epsilon$ and $\eta$ as
\begin{eqnarray}
&& \epsilon \equiv -\frac{\dot{H}}{H^2} 
 = \frac{M_P^2V'^2}{2V^2} 
 \frac{1+\frac{V}{\lambda}}{\left(1+\frac{V}{2\lambda}\right)^2}, \\
&& \eta \equiv \frac{V''}{3H^2}
 = \frac{M_P^2V''}{V}\frac{1}{1+\frac{V}{2\lambda}} \label{def-eta},
\end{eqnarray}
and the amplitude of the scalar density perturbation is given as
\begin{equation}
\mathcal{P}^{1/2}_{\zeta} = \frac{H^2}{2\pi|\dot{\phi}|}
 \simeq \frac{V^{3/2}}{\sqrt{12\pi^2}M_P^3V'}
 \left(1+\frac{V}{2\lambda}\right)^{3/2} . 
\end{equation}
This value has been normalized by COBE to be
$\mathcal{P}^{1/2}_{\zeta} = 4.7 \times 10^{-5}$. The amplitude of the scalar
density perturbation is often written as 
\begin{equation}
A_S^2 \equiv \frac{4}{25}\mathcal{P}_{\zeta}
\simeq \frac{V^3}{75\pi^2M_P^6V'^2}\left(1+\frac{V}{2\lambda}\right)^3,
\end{equation}
where we  are working under the assumption that there is no other five
dimensional impact on the inflaton other than through the modification of
Eq.~(\ref{BranefulFreidmann}) \cite{MaartensWandsBassettHeard}. 
In what follows, we adopt these formula, not least because the true five
dimensional treatment is still being developed \cite{Koyama}.
The scalar spectral index is expressed in terms of the slow roll parameters as
\begin{equation}
n_s -1 \equiv \frac{d \ln A_S^2}{d \ln k} = -6 \epsilon +2 \eta,
\end{equation}

Although we do not yet have a true five dimensional framework to discuss the
scalar perturbations, things are different for the tensor modes. In that case
the amplitude of the tensor perturbations has been obtained and is given by
\cite{LangloisMaartensWands}
\begin{equation}
\mathcal{P}_{g} = \frac{2}{M_P^2}\left(\frac{H^2}{2\pi}\right)^2 F^2(x),
\end{equation}
 or
\begin{equation}
A_T^2 \equiv \frac{1}{25}\mathcal{P}_{g}
 = \frac{2}{25M_P^2}\left(\frac{H^2}{2\pi}\right)^2 F^2(x),
\end{equation}
 with 
\begin{equation}
F^2(x) = \left[\sqrt{1+x^2}-x^2\sinh^{-1}\left(\frac{1}{x}\right)\right]^{-1},
\end{equation}
 where
\begin{equation}
x^2 = \frac{6H^2M_P^2}{\lambda}.
\end{equation}

For completeness we define the scalar-tensor ratio as
\begin{equation}
R \equiv 16\frac{A_T^2}{A_S^2} = 4\frac{\mathcal{P}_g}{\mathcal{P}_{\zeta}},
\end{equation}
 following \cite{WMAP2}.

\section{The modified gravitino bound applied to specific models}

In this section, we study several inflation models in brane world scenarios,
and constrain the allowed regions of parameter space by imposing known 
Cosmic Microwave Background radiation (CMB) anisotropy 
constraints and estimating the reheating temperature associated with the end of
inflation in the $\rho^2$ regime.

\subsection{Chaotic inflation}

The potential for chaotic inflation is given as
\begin{equation}
V(\phi) = \frac{1}{2}m^2\phi^2 .
\label{chaotic-pot}
\end{equation}
The value of the inflaton field at the end of inflation is
\begin{equation}
\phi_e = 2\sqrt{\frac{M_P}{m}}\lambda^{1/4},
\end{equation}
with the field value being related to the number of e-folds $N$ through
\begin{equation}
N \simeq \frac{m^2}{32M_P^2}\frac{\phi^4}{\lambda}
 = \frac{m^2}{192M_5^6}\phi^4. \label{efolds}
\end{equation}
The corresponding spectral index is given by
\begin{equation}
n_s-1 = -\frac{5}{2N}
\end{equation}
 and depends only on $N$.
The amplitude of the scalar density perturbations is given by
\begin{eqnarray}
\frac{H^2}{2\pi |\dot{\phi}|}
 &=& \frac{(16\times 12)^{5/4}}{144\times 8\pi}
 \left(\frac{m}{M_5}\right)^{3/2}N^{5/4} \\
 &\simeq&
 4.7\times 10^{-5}\left(\frac{m/M_5}{0.0038}\right)^{3/2}N^{5/4} \label{Qcobe}.
\end{eqnarray}
where we have deliberately written  Eq.~(\ref{Qcobe}) in terms of the COBE
result for the amplitude of the anisotropies. 
The energy scale of
inflation follows from  Eqs.~(\ref{efolds}) and (\ref{Qcobe}). From
Eq.~(\ref{Qcobe}), imposing the COBE constraint we have 
\begin{equation}
\frac{mN^{5/6}}{M_5} \simeq 0.0038
\label{mass-constraint}
\end{equation}
which can be combined with Eq.~(\ref{efolds}) in Eq.~(\ref{chaotic-pot}) to
give
\begin{equation}
V = \frac{1}{2}m^2\phi^2 \simeq \left({0.4 M_5 \over N^{1/{12}}}\right)^4 .
\label{QV}
\end{equation}
The required number of e-folds is evaluated as
\begin{equation}
N_{COBE} = 62-\ln\frac{10^{16} \textrm{GeV}}{V^{1/4}}
 +\ln\frac{V^{1/4}}{V_f^{1/4}}-\frac{1}{3}\ln\frac{V_f^{1/4}}{\rho_{rh}^{1/4}},
 \label{Ncobe}
\end{equation}
where $V$ is the potential when the COBE scales left the Hubble radius, $V_f$
is
the value at the end of inflation, and $\rho_{rh}$ is the energy density in
radiation as a result of reheating \cite{LythRiotto}. 
Substituting Eq.~(\ref{QV}) into Eq.~(\ref{Ncobe}), and dropping the negligible
third and fourth terms, we obtain
\begin{equation}
N_{COBE} = 47-\ln\frac{10^{10} \textrm{GeV}}{M_5} - \frac{1}{12} \ln N_{COBE}
\label{Ncobe2}
\end{equation}
 where we have normalized the expression in terms of $M_5 \simeq 10^{10}$ GeV
 bearing in mind the gravitino abundance constraint derived earlier 
 in Eq.~(\ref{M5bound}). 
The logarithmic dependence on $N_{COBE}$ is negligible and can be ignored,
 allowing us to estimate the mass of the inflaton in Eq.~(\ref{mass-constraint})
 as 
\begin{eqnarray}
m &\simeq& 1.5 \times 10^{-4}M_5\left(\frac{47}{N}\right)^{5/6} \\
 &=& 1.5 \times 10^6 \textrm{GeV}\left(\frac{M_5}{10^{10}\textrm{GeV}}\right)
\left(\frac{47}{N}\right)^{5/6} .
 \label{Qmass}
\end{eqnarray}

We now turn our attention to estimating the reheating temperature after 
 inflation in the $\rho^2$ dominated period. 
In the case that the inflaton decays through Yukawa interactions
 with a coupling $y$ to the matter fields, then the decay rate is approximately
 given by $\Gamma \sim y^2 m$ \cite{KolbTurner}. 
If $y = O(1) $, then the decay proceeds rapidly and reheating is almost 
instantaneous. The corresponding reheating temperature is estimated using
Eq.~(\ref{BranefulFreidmann}) with $H = \Gamma$ and Eq.~(\ref{Qmass}) to be
\begin{equation}
\label{TR-Yuk}
T_R \sim 10^9 \textrm{GeV}\left(\frac{y}{1}\right)^{1/2}
 \left(\frac{M_5}{10^{10}\textrm{GeV}}\right) ,
\end{equation}
We have already seen in Eqs.~(\ref{define-Tt}-\ref{M5bound}) that the
transition temperature where the $\rho^2$ term becomes sub-dominant is given by
 $T_t \simeq 10^6$ GeV for $M_5 = 10^{10}$ GeV. 
Since in this example $T_R \gg T_t$,  we are led to conclude 
that this brane world inspired mechanism to avoid the gravitino problem
 proposed in Ref~\cite{OkadaSeto} is workable
 for an inflation model with a quadratic potential and $y\geq 10^{-4}$.
This result is in fact applicable to many models 
 because the natural magnitude of the Yukawa coupling $y$
typically lies between $\mathcal{O}(0.1)  <y< \mathcal{O}(1) $.
This conclusion is not affected when we include the final term 
in Eq.~(\ref{Ncobe}).
 The Yukawa coupling dependence on $N$ is expressed through the $\rho_{rh}^{1/4}
\propto T_R \propto y^{1/2}$, which in Eq. (\ref{Ncobe}) leads to a change
$\Delta N$ in $N_{COBE}$ given by
 $\Delta N = (1/6)\ln y$, clearly a small change.

Of course, another possible decay route is if the inflaton decays primarily
through gravitationally suppressed interactions like a moduli field. 
In that case the decay rate is given by \cite{Coughlan:1983ci}
\begin{equation}
\Gamma \sim \frac{m^3}{M_5^2} ,
\end{equation}
and the corresponding reheating temperature is
\begin{equation}
T_R \sim 10^7 \textrm{GeV} \left(\frac{M_5}{10^{10}\textrm{GeV}}\right) .
\label{reheat-bound}
\end{equation}
This tells us that the reheating finishes during the $\rho^2$ dominant epoch
even if the inflaton decays through $M_5$ suppressed interactions. However in
this case, because $T_R$ satisfies the constraint given by
Eq.~(\ref{TR-constraint}), we do not have to rely on the new $\rho^2$ mechanism
to avoid the gravitino problem.  

\subsection{Natural inflation}

We now turn our attention to the case of natural inflation where inflation is
driven by a pseudo Nambu-Goldstone boson \cite{FreeseFriemanOlinto}. 
In what follows in this section there is significant overlap with the results
presented by Felipe \cite{Felipe}. We repeat it here, because our emphasis
is somewhat different in that we are concerned primarily with the issue of
avoiding the production of too many gravitinos. 

The potential for natural inflation is given by
\begin{equation}
V(\phi) = \Lambda^2\left[1+\cos\left(\frac{\phi}{f}\right)\right],
\label{nat-pot}
\end{equation}
 where $f$ is a spontaneous symmetry breaking scale, and $\Lambda^2$ is the
energy scale associated with inflation. Such a potential can be found in the
context of supergravity models, for example
as has been shown in \cite{MoroiTakahashi}.

The solution of the equation of motion in terms of the number of e-folds of
inflation, $\phi(N)$ is
\begin{equation}
\frac{12 M_5^6}{\Lambda^2f^2}N
 = -2\ln\frac{1-\cos(\phi/f)}{1-\cos(\phi_e/f)}-\cos(\phi/f)+\cos(\phi_e/f),
 \label{eom-N}
\end{equation}
where $\phi_e$ is the value of the inflaton at the end of inflation, 
which is in turn determined by the slow roll parameter $\epsilon=1$, 
\begin{eqnarray}
\epsilon(\phi_e)
 = \frac{12 M_5^6}{\Lambda^2f^2}\frac{1-\cos(\phi_e/f)}{(1+\cos(\phi_e/f))^2} 
 = 1.
 \label{inf-end-hybrid}
\end{eqnarray}
The spectral index is expressed exactly as
\begin{equation}
n_s-1
 = -2\frac{12 M_5^6}{\Lambda^2f^2}\frac{3-2\cos(\phi/f)}{(1+\cos(\phi/f))^2}.
\end{equation}
Using the Wilkinson Microwave Anisotropy Probe (WMAP) constraints 
on the spectral index we know that $n_s-1 \lesssim 10^{-2}$, hence we have 
\begin{equation}
\frac{12 M_5^6}{\Lambda^2f^2} \lesssim 10^{-2}.
\end{equation}
Using this constraint in Eq.~(\ref{inf-end-hybrid}), we solve the quadratic to
obtain 
\begin{equation}
\cos(\phi_e/f) \simeq -1+\sqrt{\frac{24 M_5^6}{\Lambda^2f^2}}
 \simeq -1 ,
\end{equation}
 and this allows us to write down the simplified expression for
Eq.~(\ref{eom-N}):
\begin{equation}
\frac{12 M_5^6}{\Lambda^2f^2}N \simeq \cos(\phi/f) +2\ln2-1 ,
\end{equation}
valid for $|\cos(\phi/f)| \ll 1$.

The amplitude of the density perturbation is given by
\begin{equation}
\frac{H^2}{2\pi |\dot{\phi}|}
 = \frac{1}{\pi}\left(\frac{M_5}{f}\right)^3
 \left(\frac{\Lambda^2f^2}{12 M_5^6}\right)^2
 \frac{(1+\cos(\phi/f))^3}{\sin(\phi/f)}.
\end{equation}
Using the COBE normalization, we obtain 
\begin{equation}
\frac{f}{M_5} =
 4 \times 10^2 \left(\frac{10^{-2}}{12 M_5^6/\Lambda^2f^2}\right)^{2/3},
\end{equation}
or equivalently 
\begin{equation}
\left(\frac{2\Lambda^2}{M_5^4}\right)^2
 \simeq 8.5 \times 10^{-2}\left(\frac{M_5}{f}\right)
 \simeq 2 \times 10^{-4}\left(\frac{4 \times 10^2}{f/M_5}\right).
 \label{pot-height}
\end{equation}
Thus, from  Eq.~(\ref{nat-pot}) we see that the corresponding energy scale for
natural inflation is below $M_5^4$.

The scalar-tensor ratio is given as
\begin{eqnarray}
R &=& 24 \left(\frac{12 M_5^6}{f^2\Lambda^2}\right)
 \frac{\sin(\phi/f)}{(1+\cos(\phi/f))^3} \\
 &\simeq& 0.24 \left(\frac{12 M_5^6/f^2\Lambda^2}{10^{-2}}\right)
 \qquad\textrm{for} \quad \cos(\phi/f)\sim 0,
\end{eqnarray}
 which is on the margins of acceptability. 
The mass of the inflaton field is estimated as
\begin{eqnarray}
\label{masssq}
m^2 = \frac{\Lambda^2}{f^2}
 \simeq 5 \times 10^{-8}M_5^2
 \left(\frac{12 M_5^6/f^2\Lambda^2}{10^{-2}}\right)^{5/3} ,
\end{eqnarray}
which is of the same scale as the mass of the field for the quadratic
potential.
Hence, we expect the reheating temperature would be similar.
However, the details of the allowed interactions in natural inflation models
are
different from that of polynomial inflation models where an inflaton field
simply
has the mass term and Yukawa interactions. In natural inflation, the inflaton
could decay through processes occurring at the one loop level 
\cite{MoroiTakahashi}.
For a decay rate $\Gamma \simeq 10^{-5}m^3/f^2$ in \cite{MoroiTakahashi}, 
the reheating temperature is estimated as
\begin{equation}
T_R \sim 10^5 \textrm{GeV} \left(\frac{M_5}{10^{10} \textrm{GeV}}\right)^{1/2} ,
\end{equation}
which is low enough to satisfy the bound in Eq.~(\ref{TR-constraint}), hence
does not require the new mechanism associated with $T_t$ to avoid the gravitino
problem. 
However, if the reheating process is more effective or instantaneous,
as assumed in Ref.~\cite{Felipe}, we would again be faced with a conventional
gravitino problem
for $M_5 > 10^{10}$ GeV in the supersymmetric extended models. This can be seen
by noting that the mass of the inflaton field in Eq.~(\ref{masssq}) is
comparable to that of the mass in the quadratic potential. Therefore the
maximum reheating temperature we could expect for natural inflation is comparable
to that for the quadratic inflation case which we gave in Eq.~(\ref{TR-Yuk}).
Fortunately, this is in the regime where the gravitino problem can be
alleviated using the mechanism proposed
in Ref.~\cite{OkadaSeto} if $M_5 \lesssim 10^{10}$ GeV. 

\subsection{Hybrid inflation I : D-term inflation}

Hybrid inflation models are relatively easy to find in the context of 
supersymmetry \cite{Copeland:1994vg,LythRiotto}, 
and so it is natural for us to consider them in this context through 
supersymmetric hybrid inflation.
In this subsection we will consider the case of D-term inflation
 \cite{DtermInflation}
 and in the next subsection we will turn our attention to F-term inflation. One
of the more appealing aspects of inflation models driven by D-terms is that 
they do not tend to suffer from the $\eta$ problem in a four dimensional
framework \cite{LythRiotto}, and as such this appears hopeful. The minimal
model we consider here contains three matter fields, two U(1) charged fields
$\phi_\pm$ and one neutral field $S$, which becomes the inflaton. With a
corresponding superpotential,
\begin{equation}
W = \kappa S \phi_+ \phi_- ,\label{Superpotential}
\end{equation}
where $\kappa$ is Yukawa coupling, the scalar potential is written as
\begin{equation}
V =
 e^{K(S,\phi_+,\phi_-)}\kappa^2
 \left(|S|^2|\phi_-|^2 + |S|^2|\phi_+|^2+|\phi_- \phi_+|^2+...\right)
+ \frac{g^2}{2}\left(\xi + |\phi_+|^2 - |\phi_-|^2\right)^2 ,
\label{f-termpot}
\end{equation}
 where 
\begin{equation}
K(S,\phi_+,\phi_-) = \frac{|S|^2}{M_5^2}+\frac{|\phi_+|^2}{M_5^2}
 +\frac{|\phi_-|^2}{M_5^2}
\end{equation}
 is the (minimal) K\"{a}hler potential, 
 $g$ is the gauge coupling for the U(1) fields, $\xi$ is the Fayet-Illiopoulos
(FI) term
 and the dots represent higher order terms which we do not include.
Note that in this model, the cut-off scale should be $M_5$, unlike in
four-dimensional supergravity where the cut-off scale is $M_P$.
This scalar potential, Eq.~(\ref{f-termpot}) has a local and SUSY breaking 
minimum with the non vanishing vacuum energy $V=g^2\xi^2/2 $
at $\phi_{\pm}=0$ for 
\begin{equation}
|S| > |S_c|\equiv \frac{g}{\kappa}\sqrt{\xi} , 
\end{equation}
where the mass squared of $\phi_-$ at the origin is positive as
 we will show in Eq. (\ref{mass-diff}).
It is this vacuum energy which is available to realize inflation. Since the
supersymmetry 
breaking by the non vanishing D-term generates the mass difference between
 the masses of $\phi_\pm$,
\begin{equation}
m^2_\pm = e^K\kappa^2 |S|^2 \pm g^2 \xi
\label{mass-diff}
\end{equation}
 and the masses of those fermionic partners, the potential during inflation
including radiative corrections is expressed as
\begin{equation}
V(\phi) = \frac{1}{2}g^2\xi^2\left[1+\frac{g^2}{8\pi^2}
 \left(\ln\frac{\phi^2}{\Lambda^2}+\frac{\phi^2}{2M_5^2}\right)\right],
 \label{Dpotential}
\end{equation}
 where $\phi =\sqrt{2}|S|$ is the canonically normalized inflaton
 and $\Lambda$ is the renormalization scale
 \cite{DtermInflation}.
Here, the third term comes from the $e^K$ pre-factor in Eq.~(\ref{mass-diff}).
The slow roll equation of motion is
\begin{equation}
3H\dot{\phi}+ \frac{1}{2}g^2\xi^2\frac{g^2}{8\pi^2}
 \left(\frac{2}{\phi}+\frac{\phi}{M_5^2}\right)=0.
 \label{d-term-eom}
\end{equation}
Now, it is useful to divide this into two parts depending on which of the two
potential related terms in Eq.~(\ref{d-term-eom}) is dominating, namely $\phi
\lesssim \sqrt{2}M_5$ (the second stage) and $\sqrt{2}M_5 \lesssim \phi$ (the
first stage). Considering the second stage, the solution of
Eq.~(\ref{d-term-eom}) is
\begin{equation}
\frac{\phi^2}{2}
 =
M_5^2\left(\frac{M_5}{\xi^{1/2}}\right)^4\frac{6N}{\pi^2}+\frac{\phi_e^2}{2},
 \label{DsecondSolution}
\end{equation}
 where $\phi_e^2$ is either $\phi_f^2 = 6M_5^6/(\pi^2\xi^2)$ 
 or $\phi_c^2 = 2g^2\xi/\kappa^2$.
The former $\phi_f$ corresponds to the condition $\eta =-1$ and 
 the latter $\phi_c$ to the symmetry breaking point.
Since the situation for inflation only gets more difficult for the latter
condition, we continue here to consider the former case, in other words, we
consider the case where $2g^2\xi/\kappa < 6M_5^6/(\pi^2\xi^2)$. 
Now, in this model, we consider $\xi\leq M_5^2$,
 because it seems impossible that the magnitude of the FI term is larger than 
 the (fundamental) cut-off scale 
 when the FI term is induced in supergravity \cite{DvaliKallosh}. 
Given the bound, and the fact that in this second stage, we are in the regime
where $\phi \lesssim \sqrt{2}M_5$, it then follows from Eq.
(\ref{DsecondSolution}), that it is impossible to have more than
$\mathcal{O}(1)$ e-folds of
inflation during the $\rho^2$ dominated epoch. Given that disappointing
outcome, we turn our attention to the first stage. Again though we are
disappointed, the slow roll parameter $\eta$ for the first stage is given by
\begin{equation}
\eta = \frac{3}{\pi^2}\left(\frac{M_5}{\xi^{1/2}}\right)^4,
\end{equation}
which is always greater than unity unless $M_5 \sim \xi^{1/2}$, implying that
again it is very difficult to obtain sufficient e-folds of inflation. Another
way of seeing this is by considering the solution of the equation of motion
during the first stage
\begin{eqnarray}
\frac{\phi}{\sqrt{2}M_5}
 &=& e^{{\frac{3N}{\pi^2}\left(\frac{M_5}{\xi^{1/2}}\right)^4}} \\
 &\simeq& 4\times 10^6 \quad \textrm{for} 
 \quad (N, \xi^{1/2}) =(50, M_5) \label{Dinitial}. 
\end{eqnarray}
Inserting this into the condition for false vacuum energy domination
\footnote{If this condition is not satisfied, 
this model is equivalent to chaotic inflation model 
except for the phase transition after inflation.} in Eq.~(\ref{Dpotential}),
\begin{equation}
\frac{g^2}{8\pi^2}\frac{\phi^2}{2M_5^2} \ll 1,
\end{equation}
we obtain a bound on the allowed value of $g^2$
\begin{eqnarray}
\frac{g^2}{8\pi^2}
 &\ll& e^{-\frac{3N}{\pi^2}\left(\frac{M_5}{\xi^{1/2}}\right)^4} \\
 && \simeq 2.5\times 10^{-7} \quad \textrm{for} \quad (N, \xi^{1/2}) =(50,
M_5).
\end{eqnarray}
In particular we see that if we take $N \sim 50$,
 the required gauge coupling $g$ is too small to be physically sensible. 
 In addition, the necessary initial value of the inflaton is also 
too large as can be seen in Eq.~(\ref{Dinitial}) and it is even worse than 
 the case of the quadratic potential, $\phi_{COBE} \sim 10^2 M_5$.
For such a large field value region, the validity of the potential is doubtful,
 in other words, nothing guarantees the flatness of the
 inflaton potential to be preserved because of possible  nonrenormalizable terms
\cite{LythRiotto}.
Moreover, even if one accepts these facts,
 the amplitude of the density perturbations is estimated as
\begin{eqnarray}
\frac{H^2}{2\pi|\dot{\phi}|}
 &=& \frac{\pi}{72\sqrt{2}}g^2
 e^{-\frac{3N}{\pi^2}\left(\frac{M_5}{\xi^{1/2}}\right)^4}
 \left(\frac{\xi^{1/2}}{M_5}\right)^8 \nonumber \\
 &=& 7.7 \times 10^{-15} 
 \left(\frac{g^2}{10^{-6}}\right),
\end{eqnarray}
demonstrating that there is no consistent parameter. In summary, we see that
D-term inflation during a $\rho^2$ dominant era in a RS brane world type
cosmology is simply not viable, a conclusion which differs from a recent study
in Ref.~\cite{Panotopoulos}, where the discussion there was undertaken assuming
a globally supersymmetric potential with the third term 
in Eq.~(\ref{Dpotential}) being absent.
In principle, a possible way out of this problem is to consider the case when inflation occurs 
after the $\rho^2$ term has become negligible, a situation which correspond to models with 
a high mass $M_5$, or in other words, a high brane tension. Unfortunately, even for this case,
we still obtain the severe lower bound $ M_5 > \mathcal{O}(0.1) M_P$ for 
$g^2 =\mathcal{O}(1)$ from
\begin{equation} 
\eta = 
\frac{g^2}{8\pi^2}M_P^2\left(-\frac{2}{\phi^2}+\frac{1}{M_5^2}\right) \ll 1,
\end{equation}
which in turn is obtained from Eq. (\ref{Dpotential}) 
and Eq. (\ref{def-eta}) for $2\lambda \gg V$.
Hence, for all practical purposes we find that $M_5 \simeq M_P$ is required.

\subsection{Hybrid inflation II : F-term inflation}

We now turn our attention to the case hybrid inflation driven by 
F-terms \cite{Copeland:1994vg,LythRiotto}.
The superpotential for these class of hybrid inflation models is given by
\begin{equation}
W = \kappa S\bar{\Psi}\Psi- S \mu^2 ,
\end{equation}
 where $S$ is a gauge singlet superfield, $\Psi$ and $\bar{\Psi}$ are a
conjugate pair of superfields
 transforming as a nontrivial representation of some gauge group
\cite{DvaliShafi}.
The scalar potential in the global supersymmetry limit is expressed as
\begin{equation}
V = (\kappa |\Psi|^2-\mu^2)^2 + 2 \kappa^2|S|^2|\Psi|^2,
\end{equation}
where the D-flat condition $\Psi^* = \bar{\Psi}$ is imposed.
For $ |S| > S_c \equiv \mu/\sqrt{\kappa}$, the potential is minimized at
$\Psi=0$
 and has the false vacuum energy $\mu^4$.
The potential including 1-loop radiative corrections 
\cite{DvaliShafi} and supergravity corrections is given by
\begin{equation}
V(\phi) = \mu^4\left[1+\frac{\kappa^2}{16}\ln\frac{\phi^2}{\Lambda^2}
 +c_1\frac{\phi^2}{M_5^2}+c_2\frac{\phi^4}{M_5^4} +... \right] ,
 \label{Fpotential}
\end{equation}
 where $\phi=\sqrt{2}|S|$ is the canonically normalized inflaton,
 $c_1$ and $c_2$ are constants which come from the K\"{a}hler potential 
 and would normally be order of unity. 
The corresponding slow roll equation is written as
\begin{equation}
3H\dot{\phi} +\mu^4\left[\frac{\kappa^2}{16}\frac{2}{\phi}
 +2c_1\frac{\phi}{M_5^2}+4c_2\frac{\phi^3}{M_5^4} +...\right] =0 ,
\end{equation}
which we can again divide into two parts depending upon which terms are
dominating.  
The first stage is where the potential is dominated by the polynomial terms and
the second stage corresponds to case where the logarithmic term dominates. 
The slow roll parameter $\eta$ for the potential (\ref{Fpotential})
 is expressed as
\begin{equation}
\eta = \left(-\frac{3 \kappa^2 M_5^2}{2\phi^2}+24c_1
 +144 c_2\left(\frac{\phi}{M_5}\right)^2\right) \left(\frac{M_5}{\mu}\right)^4
,
\end{equation}
and on the face of it, it looks as if the $\eta$ problem will generally be
present, because the constant terms $c_1$ and $c_2$ are $\mathcal{O}(1)$ and
$\mu \leq M_5$.

To be more specific we begin the investigation by considering the inflationary
dynamics during the second stage described above, which is relevant for values
of $\phi$ satisfying 
\begin{equation}
\frac{\phi^2}{M_5^2} < \frac{\kappa^2}{16 c_1}.
\label{SecondRegion}
\end{equation}
The solution of the equation of motion is given as
\begin{equation}
\frac{\phi^2}{M_5^2}
 = 3\kappa^2 N \left(\frac{M_5}{\mu}\right)^4+\frac{\phi^2_e}{M_5^2},
 \label{FSecondSolution}
\end{equation}
 where $\phi_e^2$ is either 
\begin{equation}
\phi_f^2 = \frac{3\kappa^2}{2}M_5^2\left(\frac{M_5}{\mu}\right)^4,
\end{equation}
 which corresponds to $\eta=-1$, or
\begin{equation}
\phi_c^2 = \frac{2\mu^2}{\kappa},
\end{equation}
which corresponds to the symmetry breaking point.
Combining Eqs.~(\ref{SecondRegion}) and (\ref{FSecondSolution}) we can write
\begin{equation}
c_1 < \frac{1}{48 N}\left(\frac{\mu}{M_5}\right)^4
 = 4.2 \times 10^{-4}\left(\frac{50}{N}\right)\left(\frac{\mu}{M_5}\right)^4 ,
\end{equation}
 for $\phi_e=\phi_f$. 
In other words $c_1$ must be extremely small if we are to have sufficient
numbers of e-folds during the second stage defined by Eq.~(\ref{SecondRegion}). 
This fine tuning is a specific example of the more general $\eta$ problem, we
referred to earlier. 
The constraint becomes more severe for $\phi_e=\phi_c$, so from now on we will
only consider the case of $c_1=0$. Although this is a very specific selection
of coefficients, it is in fact possible to construct such a model in
supergravity
through a combination of accidental cancellations, if the K\"{a}hler potential
is minimal. 
However, in that case, we find that $c_2$ must satisfy $c_2 = 1/8$
\cite{LindeRiotto}.

Returning to the second stage of inflation with $c_1=0$, 
the new relevant region for the field is 
\begin{equation}
\frac{\phi^4}{M_5^4} < \frac{\kappa^2}{4}\left(\frac{1/8}{c_2}\right) ,
\label{SecondRegion2}
\end{equation}
 instead of Eq.~(\ref{SecondRegion}).
We are not going to consider the case of $\phi_e = \phi_c (> \phi_f)$, because
it is not possible in that case to simultaneously satisfy the inequalities
(\ref{SecondRegion2}) and $\phi_c > \phi_f$ while maintaining the condition 
$\mu \leq M_5$.
For the case of $\phi_e = \phi_f$, Eq.~(\ref{SecondRegion2}) is rewritten as
\begin{equation}
\kappa < 
\frac{1}{6N}\left(\frac{\mu}{M_5}\right)^4\left(\frac{1/8}{c_2}\right)^{1/2} .
\label{SecondRegion2'}
\end{equation}
On the other hand, the condition $\phi_f > \phi_c$ is equivalent to 
\begin{equation}
\kappa^3 > \frac{4}{3}\left(\frac{\mu}{M_5}\right)^6 .
\label{FLargerThanC}
\end{equation}
Both Eqs.~(\ref{SecondRegion2'}) and (\ref{FLargerThanC}) are satisfied
simultaneously for $\mu \leq M_5$ only when
\begin{equation}
N < \frac{1}{288^{1/3}} \left(\frac{1/8}{c_2}\right)^{1/2} \simeq 0.15.
 \label{Nbound}
\end{equation}
for $c_2 \sim 1/8$. It follows that in order to obtain a realistic number of
e-folds, say $N=50$, then $c_2 \lesssim 10^{-6}$, another case of severe fine
tuning.  Thus, we can say that in general it is impossible to obtain enough
expansion during the second stage even if $c_1 =0$.

Next, we consider the first stage of inflation driven by the polynomial terms
in the potential Eq.~(\ref{Fpotential}). 
There we have
\begin{equation}
\frac{\kappa^2}{32}
 < \frac{1}{8}\frac{\phi^4}{M_5^4}\left(\frac{c_2}{1/8}\right) \ll 1 ,
 \label{FirstRegion}
\end{equation}
 where the upper bound comes from the condition for false vacuum domination 
 and the lower bound comes from Eq.~(\ref{SecondRegion2}).
The solution of the slow roll equation for this first stage, 
\begin{equation}
3H\dot{\phi}+\mu^4 4c_2\frac{\phi^3}{M_5^4}=0 ,
\end{equation}
 is given by 
\begin{equation}
-\frac{M_5^2}{2\phi^2}+\frac{M_5^2}{2\phi^2_*}
 =6 \left(\frac{c_2}{1/8}\right)\left(\frac{M_5}{\mu}\right)^4(N-N_*) ,
 \label{FirstSolution}
\end{equation}
 where
\begin{equation}
\phi_*^2= \frac{\kappa}{2}M_5^2 \left(\frac{1/8}{c_2}\right)^{1/2} ,
\end{equation}
arises from the lower bound in Eq.~(\ref{FirstRegion}) and
\begin{equation}
N_* = \frac{1}{6\kappa}\left(\frac{\mu}{M_5}\right)^4
\left(\frac{c_2}{1/8}\right)^{1/2},
\label{N_*bound}
\end{equation}
 for $\phi_e = \phi_f > \phi_c$. 
Again, $\phi_c > \phi_f$ is incompatible with Eq.~(\ref{FirstRegion}).
Equation~(\ref{FirstSolution}) can be rewritten as 
\begin{eqnarray}
\frac{M_5^2}{\phi^2}
 = \frac{4}{\kappa}\left(1-3\kappa\left(\frac{M_5}{\mu}\right)^4 N\right)
 = \frac{4}{\kappa}\left(1-\frac{N}{2N_*}\right) .
\end{eqnarray}
We now see that the total number of possible e-folds $N$ in this first stage is
constrained by $N < 2N_*$.
Since $N_*$ is bounded by Eq.~(\ref{N_*bound}), 
with Eq.~(\ref{FLargerThanC}) we obtain 
\begin{equation}
N \leq \mathcal{O}(0.1) .
\end{equation}
Thus, we conclude that it is impossible to realize F-term hybrid inflation
 during a $\rho^2$ dominated era in a RS brane world cosmology. This result is
similar to that obtained in the case of inflation models in M-theory by Lyth
\cite{lyth-non-renorm}
In addition, unlike D-term inflation model, even if we take $M_5=M_P$,
these models still have the famous $\eta$ problem, 
and this difficulty can not be removed.

\section{Inflation models with steep exponential potentials}

Inflation arising in $\rho^2$ cosmologies was investigated in detail for the
case of a steep exponential potential in Ref.~\cite{SteepInflation}. The 
potential can be written as
\begin{equation}
V(\phi) = V_0 e^{-\alpha \phi/M},
\end{equation}
where $V_0$ and $\alpha$ are constants and $M$ is a large mass scale.
The slow roll parameters in this model are 
\begin{eqnarray}
\epsilon = \eta = \frac{12M_5^4}{V}\left(\frac{\alpha M_5}{M}\right)^2 ,
\end{eqnarray}
and the end of inflation is given by $\epsilon = \eta =1$.
These models are interesting because of this ability to naturally terminate
inflation, a property not available to exponential potentials in conventional
Einstein gravity.  
In particular, the number of e-folds is written as
\begin{equation}
V = 12 (N+1) M_5^4 \left(\frac{\alpha M_5}{M}\right)^2 .
\label{SteepVNrelation}
\end{equation}

The condition to solve the horizon problem is expressed as
\begin{equation}
H_{inf}^{-1}e^N\left(\frac{a_0}{a_{end}}\right) \gtrsim H_0^{-1}
\label{SteepHorizon}
\end{equation}
 where $H_{inf}$ is the Hubble parameter during inflation, $a_{end}$ is
 the scale factor at the end of inflation and the subscript $0$ denotes
 the present value.
One particular characteristic of this type of steep inflation model is that 
radiation is generated 
not by the decay of the inflaton but through gravitational particle production.
Since the energy density of the particles produced at the end of inflation is
estimated as
 $\rho_{rad} \simeq T_{end}^4 \sim H_{end}^4$ \cite{GraviProduction},
 we obtain
\begin{equation}
\frac{a_0}{a_{end}} = \frac{T_{end}}{T_0} \sim \frac{H_{end}}{T_0} .
\label{Tend}
\end{equation}
From Eqs.~(\ref{SteepVNrelation}) - (\ref{Tend}) and (\ref{BranefulFreidmann}),
 we find
\begin{equation}
\frac{e^N}{N+1} \gtrsim \frac{T_0}{H_0}
 \simeq 1.6\times 10^{29}\left(\frac{0.7}{h}\right) ,
\end{equation}
which implies 
\begin{equation}
N \gtrsim 72.
\end{equation}
This number of e-folds corresponds to the present horizon scale and is 
larger value than that initially estimated in the original paper
\cite{SteepInflation}.
It follows that the density perturbation and the COBE normalization are given
as
\begin{eqnarray}
\frac{H^2}{2\pi|\dot{\phi}|}
 &=& \frac{(N+1)^2}{\pi}\left(\alpha\frac{M_5}{M}\right)^3 \nonumber \\
 &=& 4.7 \times 10^{-5}\left(\frac{N+1}{73}\right)^2
 \left(\frac{(\alpha M_5/M)^3}{2.8 \times 10^{-8}}\right),
\end{eqnarray}
which is consistent for $\frac{(\alpha M_5/M)^3}{2.8 \times 10^{-8}}\sim 1$.
The spectral index of the density perturbation is revised upwards
as
\begin{equation}
n_S =1- \frac{4}{N+1} \simeq 0.94 \quad \textrm{for} \quad N \simeq 72 ,
\end{equation}
and the scalar-tensor ratio is revised downward as,
\begin{eqnarray}
R = \frac{24}{N+1} = 0.33 \left(\frac{N+1}{73}\right). 
\end{eqnarray}
These numbers are marginally consistent with the allowed region of parameter
space arising from the first year WMAP data \cite{WMAP}.
However, when we compare with the joint likelihood data analysis using both
WMAP
and other observational results on the large scale structure such as 
 Sloan Digital Sky Survey (SDSS), 
we find that it is still excluded at the $3\sigma$ level
\cite{TsujikawaLiddle}.
A possible resolution of this problem involving the curvaton scenario
\cite{Curvaton} was recently proposed by Liddle and Urena-Lopez
\cite{LiddleCurvaton}. 
For now we continue to investigate the
production of gravitinos based on the original model.
Since the curvaton field can be an additional source of 
late-time entropy production and dilute the gravitino abundance,
if the original model does not have the gravitino problem,
then steep inflation with a curvaton would also avoid it.

Now, to consider gravitino production, we need to determine the evolution of the
universe after the inflationary period. It is important to realize that the
thermal history in this particular model is different from that assumed earlier
in that the production of radiation comes about here from gravitational
interactions and is less efficient than in the conventional cases considered
earlier. This means that we can no longer use the constraints from
Eqs.~(\ref{Tt}) and (\ref{M5bound}) in the case of the steep exponential
potentials. They do not apply. 
The energy densities of the inflaton and gravitationally produced particles
 are given by
\begin{eqnarray}
\left.\rho_{\phi}\right|_{end} &=& 2V_{end} 
 = 48M_5^4\left(\frac{\alpha M_5}{M}\right)^2 , \\
\left.\rho_{rad}\right|_{end} &\sim & H_{end}^4 
 = 16M_5^4\left(\frac{\alpha M_5}{M}\right)^8 ,
\end{eqnarray}
 respectively.
However, gravitons (gravitational
waves with a short wavelength) are also produced simultaneously, a fact that
leads to another problem 
 for the original steep inflation model \cite{GravitonProblem}.
Of course, if late time entropy production occurs, 
we can overcome this problem too \cite{LiddleCurvaton}.
After the steep inflation period terminated, the energy density of the inflaton
is dominated
 by its kinetic energy. 
This is the kination epoch of the scalar field with an exponential
 potential, where the energy density of the scalar field decreases as
 $\rho_{\phi} \propto a^{-6}$, while that of radiation decreases 
 as $\rho_{rad} \propto a^{-4}$.
The time of inflaton-radiation equality is therefore determined by 
\begin{equation}
\left(\frac{a_{end}}{a_{\phi - rad}}\right)^2
 = \frac{1}{3}\left(\frac{\alpha M_5}{M}\right)^6 ,
\end{equation}
 where $a_{\phi - rad}$ is the scale factor at equality.
The corresponding energy density of the radiation is estimated as
\begin{eqnarray}
\left.\rho_{rad}\right|_{\phi-rad} &=& \left.\rho_{rad}\right|_{end}
\left({a_{end} \over a_{\phi-rad}}\right)^4 \nonumber \\
 &=& \frac{16}{9}M_5^4 \left(\frac{\alpha M_5}{M}\right)^{20} =
\frac{\pi^2}{30}g_* T_{eq}^4
  \label{DefTeq} \\
 &\simeq& (2.0 \times 10^{-13} M_5)^4
 \left(\frac{\alpha M_5/M}{2.8 \times 10^{-3}}\right)^{20} .
 \label{define-Teq}
\end{eqnarray}
This radiation must be the dominant contribution to the total energy density by
the time of BBN, which corresponds to a temperature of order 1 MeV. Hence we
obtain the constraint (for $\frac{\alpha M_5/M}{2.8 \times 10^{-3}} \sim 1$), 
\begin{equation}
M_5 \gg 5\times 10^9 \textrm{GeV}.
\end{equation}
However, 
this is the most conservative constraint on $M_5$. 
If we insist that we do not overproduce neutralinos as the LSP, then 
we obtain the tighter constraint
\begin{equation}
M_5 \gtrsim 2.5\times 10^{14} \textrm{GeV}
 \left(\frac{m_{\chi}}{100 \textrm{GeV}}\right),
 \label{LSPconst}
\end{equation}
 where $m_{\chi}$ denotes the mass of the neutralino
\footnote{
We will shortly show, that
 the abundance of gravitinos in this model is very small, thus 
 provided the LSP forms dark matter, then 
 the neutralino is the most promising candidate. 
In addition, this constraint has not been considered in the previous section,
 because this condition gives the lower bound $M_5 \gtrsim 10^5$ GeV
 for the neurtalino dark matter \cite{OkadaSetoDM} and
 is just consistent with Eq.~(\ref{M5bound}) 
 in the standard case where the reheating finishes by the inflaton decay.
 }. 
In this simple scenario, 
 the relic number density of dark matter particles freezes out 
 when the annihilation rate $\Gamma_{ann}$ becomes smaller than the cosmic
expansion rate,
\begin{equation}
\Gamma_{ann} < H.
\end{equation}
The bound Eq.~(\ref{LSPconst}) arises 
 because the additional energy source coming from the inflaton, $\rho_{\phi}$, 
 increases $H$, leading to an earlier time of decoupling of
 the neutralino and consequently an enhancement of the relic density. 

Since the brane tension $\lambda$ (or the transition temperature $T_t$) is
given
by Eq.~(\ref{define-Tt}), 
\begin{eqnarray}
2\lambda &=& \frac{12M_5^6}{M_P^2} = \frac{\pi^2}{30}g_* T_t^4 \nonumber \\
 &\simeq& 2\times 10^{-18} M_5^4 \left(\frac{M_5}{10^9\textrm{GeV}}\right)^2 ,
 \label{define-Tt2}
\end{eqnarray}
 we find by comparing Eq.~(\ref{define-Tt2}) with Eq.~(\ref{define-Teq}) that 
$T_t \gg T_{eq}$, a point that we will make use of shortly.

Finally, we estimate the abundance of the gravitinos produced at the end of the
period of inflation and show that in the context of steep inflation it is not a
problem.
The relevant Boltzmann equation for the gravitino production 
is Eq.~(\ref{GravitinoBoltamann}). This needs to
be combined with the modified Friedmann equation Eq.~(\ref{BranefulFreidmann}),
recalling that the total energy density is given by $\rho = \rho_\phi +
\rho_{rad}$. Making use of Eq.~(\ref{M_5-M_P}) we have
\begin{equation}
H= \sqrt{{\rho_{rad} \over 3M_P^2}}  \sqrt{1 + {\rho_\phi \over \rho_{rad}}}
\sqrt{1 + {M_P^2 \over 12M_5^6} (\rho_\phi + \rho_{rad})}, 
\end{equation}
hence Eq.~(\ref{GravitinoBoltamann}) becomes:
\begin{equation}
\frac{d Y_{3/2}}{d T} = 
 -\frac{s\langle\sigma_{tot} v\rangle}
 {\sqrt{\frac{\rho_{rad}}{3M_P^2}}T}Y_{rad}^2
\frac{1}{\sqrt{\left(1+\frac{\rho_{\phi}}{\rho_{rad}}\right)\left(1+\frac{M_P^2(
\rho_{\phi}+\rho_{rad})}{12M_5^6}\right)}}.
 \label{fullboltz}
\end{equation}
Here there are two effects arising from the Hubble parameter, which combine to
give an additional suppression to the overall gravitino abundance, namely,  the
$\rho^2$ term and the remaining energy density of the inflaton. The latter is
more effective than the former in this scenario because $T_t \gg T_{eq}$, 
so in what
follows we drop the $\rho_{rad}$ term in the expression $\sqrt{1 + {M_P^2 \over
12M_5^6} (\rho_\phi + \rho_{rad})}$ in Eq.~(\ref{fullboltz}).
The final gravitino abundance is obtained by integrating the Boltzmann equation
 from the reheating temperature $T_R$ to a low temperature $T_{low}$. 
Since the pre-factor
 $s\langle\sigma_{tot} v\rangle Y_{rad}^2/ \sqrt{\rho_{rad}/3M_P^2}T $ 
 is the same as in the standard case and almost constant, we concentrate on the
integration of the final part in Eq.~(\ref{fullboltz}).
The integration can be split into various regimes in which different terms are
important. For example we can imagine the system cooling through the range $T_R
\to T_t \to T_{eq} \to T_{low}$. Using $\rho_{\phi}/\rho_{rad} =(T/T_{eq})^2$,
along with Eqs.~(\ref{define-Tt}) and (\ref{DefTeq}), 
the integration can be rewritten as
\begin{eqnarray}
&& - \int_{T_R}^{T_{low}}  
 \frac{d T}{\sqrt{(1+...)(1+...)}} \nonumber \\
&& \simeq - \int_{T_{eq}}^{T_{low}} d T - \int_{T_t}^{T_{eq}} \frac{d
T}{\sqrt{\left(\frac{T}{T_{eq}}\right)^2}}- \int_{T_R}^{T_t} \frac{d
T}{\sqrt{\left(\frac{T}{T_{eq}}\right)^2}\sqrt{\frac{M_P^2}{12M_5^6}\frac{\pi^2}
{30}g_*T_{eq}^4\left(\frac{T}{T_{eq}}\right)^6}}
\\
&& \simeq
 T_{eq}\left(1+\ln\frac{T_t}{T_{eq}}+\mathcal{O}(\frac{T_{eq}}{T_t})\right)
 \sim 10 T_{eq},  
\label{BraneBoltz:Yapp}
\end{eqnarray}
hence it is characterized by $T_{eq}$.
Now since
\begin{eqnarray}
T_{eq} \sim 10^{-13} M_5
 &\simeq& \textrm{MeV}\left(\frac{M_5}{10^{10} \textrm{GeV}}\right) ,\\
 &\simeq& 100 \textrm{GeV}\left(\frac{M_5}{10^{15} \textrm{GeV}}\right) ,
\end{eqnarray}
then $10 T_{eq}$, 
 is much lower than the upper bound arising from the gravitino problem, $10^6$
GeV. Consequently 
 there is no gravitino problem with $\rho^2$ inflation driven by a steep
exponential potential.

\section{Conclusions}

In this paper we have investigated inflation driven 
by the inflaton on the brane during the $\rho^2$ term dominated era 
in a class of brane world inspired models. In particular we have 
analyzed the dynamics of a wide class of inflationary potentials, estimated the
associated reheating temperature and gravitino abundance in these models, always
taking into account constraints arising from observations of 
the CMB anisotropies by COBE and WMAP. 

Single field models we study include chaotic inflation 
 with a quadratic potential and natural inflation. Interestingly, 
for a significant range of parameters, 
these models can be made to avoid the usual gravitino problem 
in a manner first pointed out in Ref.~\cite{OkadaSeto}. Moreover,
we found that $M_5 < 10^{10}$ GeV as given in Eq.~(\ref{M5bound}) is required 
even if the inflaton decays only through gravitational interactions. 

Inflation driven by an exponential potential in a brane world cosmology
 is interesting for a number of reasons. Inflation can occur for steep
potentials, it terminates of its own accord unlike in the usual Friedmann
scenario in four dimensions, and gravitational particle production is available
as a mechanism to reheat the Universe. 
Unfortunately, there are a number of tight constraints
emerging on this particular original model \cite{TsujikawaLiddle}, and 
although in this paper we have shown how the gravitino bound can be alleviated,
the model is still on the borderline of being ruled out, requiring an additional
feature such as a curvaton type coupling to make it viable
\cite{LiddleCurvaton}.

The most telling result, clearly of significance to those 
interested in building models of inflation, 
concerns the viability of hybrid inflation
models, which arise naturally in supersymmetric theories. Since the reheating
temperature of hybrid inflation can be high in many cases, these are natural
models for us to consider given our interest in overcoming the gravitino
problem, and the fact that we are thinking of inflation occurring in high
energy regimes.
Unfortunately, the gravitino question is somewhat irrelevant here because we
have seen that these supersymmetric hybrid inflation models do not
work within $\rho^2$ inflation in RS type brane world models, a result
similar to that obtained earlier in the context of M-theory by Lyth
\cite{lyth-non-renorm}.
Even D-tem inflation is plagued with the $\eta$ problem, making it very
difficult to obtain sufficient expansion, or generate the correct density
perturbation spectrum for a sensible range of masses and coupling constants.
Similarly, we found it to be impossible for F-term inflation to generate 
sufficient expansion.
The main reason for this problem is the reduction of the cut-off scale arising
from the supergravity correction.
The difference between our result and previous studies comes from this fact, 
since these effects have not been taken into account appropriately in the
previous studies \cite{TsujikawaLiddle, Panotopoulos, BoutalebJ}. This fact
may imply that it is in general difficult to construct a viable inflation model
in the theory with a low gravitational scale. 

%
\section*{Acknowledgments}

O.S. would like to acknowledge the support of PPARC and is grateful to Andrew
Liddle and David Lyth for valuable discussions. 
We thank the Yukawa Institute for Theoretical Physics at Kyoto University,
 where this work was initiated during the YITP-W-04-13  on ``The 14th Workshop
on General Relativity and Gravitation".



\end{document}